\documentclass[prb,twocolumn,showpacs,amsmath,amssymb]{revtex4-1}

\usepackage{graphicx}

\newcommand{\superscript}[1]{\ensuremath{^{\textrm{#1}}}}

\begin{document}
\title{Disorder-induced gap behavior in graphene nanoribbons}
\author{Patrick Gallagher}
\author{Kathryn Todd}
\author{David Goldhaber-Gordon}
\email{goldhaber-gordon@stanford.edu}
\affiliation{Department of Physics, Stanford University, Stanford, California, 94305, USA}

\begin{abstract}

We study the transport properties of graphene nanoribbons of
standardized 30 nm width and
varying lengths. We find that the extent of the gap observed in
transport as a function of Fermi energy in these ribbons (the
``transport gap'') does not have a strong dependence on ribbon length,
while the extent of the gap as a function of
source-drain voltage (the ``source-drain gap'') increases with
increasing ribbon length. We anneal the ribbons
to reduce the amplitude of the disorder potential, and find
that the transport gap both shrinks and moves closer to zero gate
voltage. In contrast, annealing does not systematically
affect the source-drain gap. We conclude that the transport gap reflects
the overall strength of the background disorder potential, while the
source-drain gap is sensitively
dependent on its details. Our results support the model that transport
in graphene nanoribbons
occurs through quantum dots forming along the ribbon due to a disorder
potential induced by charged
impurities.
\end{abstract}


\maketitle

Graphene is a two-dimensional sheet of carbon atoms in which
low-energy
charge carriers obey a linear dispersion relation with no
bandgap.\cite{CastroNeto2009} However, when graphene sheets are
etched into 
``nanoribbons,'' strips of graphene of nanometer-scale widths, they
can exhibit 
gapped transport behavior: conductance can be suppressed over a range
of Fermi 
energies and source-drain biases.\cite{Han2007, Chen2007} While
theoretical models \cite{Brey2006, Ezawa2006, Yang2007}
predict that a gap in the band structure 
can open up for nanoribbons of certain edge orientations,
these models are of limited applicability 
to samples that have been studied experimentally, as 
lithographically-produced nanoribbons have edge roughness on the order
of nanometers, and the crystallographic orientation of these
nanoribbons appears not to impact transport
properties.\cite{Han2007} There is also mounting experimental evidence
that the 
gap-like behavior has a strong length
dependence,\cite{Todd2009,Molitor2009} suggesting that 
mechanisms other than a bandgap are important in the conductance
suppression. 

Several alternative mechanisms have been
proposed to explain the gap-like behavior. One 
common proposal is based on Anderson localization: calculations
based on non-interacting electrons\cite{Evaldsson2008,Mucciolo2009}  
 have shown that given a 
small amount of edge roughness or other short-range
disorder,
Anderson localization can lead to an 
appreciable ``transport gap,'' or a
region of Fermi energy around the charge neutrality point where
conductance is 
strongly suppressed at zero bias even in the absence of a band gap. 
However, recent observations of Coulomb diamond-like features in
device conductance suggest an alternate model.\cite{Todd2009,
  Stampfer2009, Molitor2009, Liu2009} For shorter ribbons,
these diamonds can be clearly resolved, and near the charge neutrality
point the
pattern of diamonds resembles that of a few quantum dots in parallel
and/or in series. For longer ribbons, the diamonds of suppressed
conductance start overlapping, as expected for multiple dots in
series. These observations have led to the proposal that
charge transport occurs mainly through an arrangement
of quantum dots along the ribbon. It has been suggested that 
quantum dots form either as a result of lithographic line-edge
roughness,\cite{Sols2007} or as a result of potential inhomogeneities
due to charged 
impurities near the ribbon, coupled with a smaller confinement-induced
energy gap between 
electrons and holes that creates ``tunnel
barrier'' regions of
zero charge carrier density between puddles. \cite{Todd2009,
  Stampfer2009}

We note that long-range scattering, as from charged impurities in the
vicinity of the graphene, is predicted not to cause Anderson
localization in extended graphene sheets,\cite{Nomura2007, Nomura20072,
  Bardarson2007} while short-range scatterers such as lattice defects
are expected to contribute to Anderson localization. Furthermore,
calculations \cite{Mucciolo2009} have indicated that long-range
scatterers in the vicinity of the graphene sheet do not cause
localization even in nanoribbon geometries. Therefore we suggest that
measurements that distinguish between the effects of lattice defects
at the ribbon edges and charged impurities in the ribbon's vicinity,
such as the annealing studies described in this paper, may also
distinguish between models of transport based on Anderson localization
or quantum dot formation.

\begin{figure}
\centering
\includegraphics[width=3.3in]{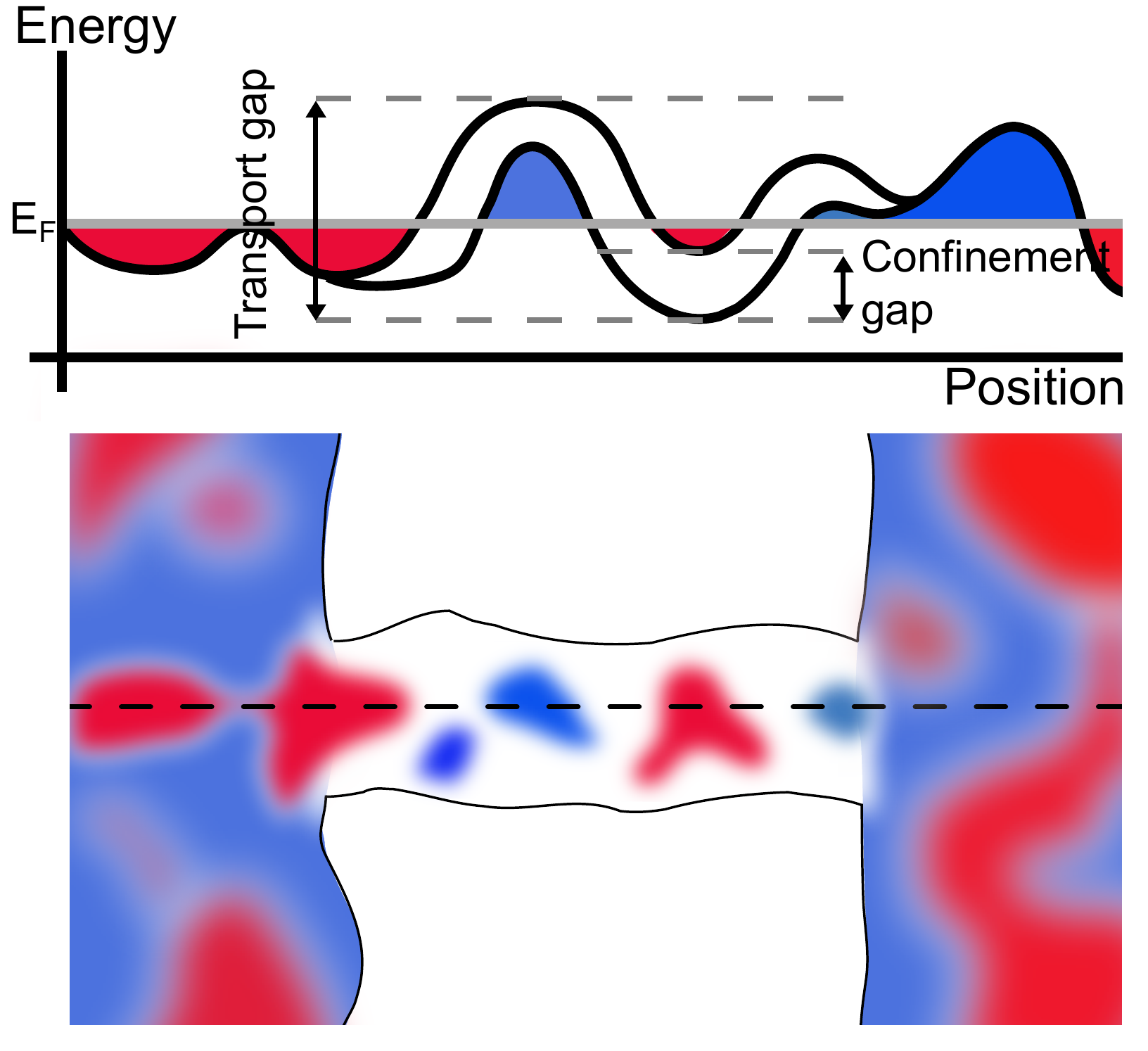}
\caption{(Color online) Cartoon of quantum dots forming along the
  ribbon due to 
  potential inhomogeneities and a confinement gap. The red (dark gray)
  puddles
  indicate electrons, and the blue (light gray) puddles 
  indicate holes. The thick dark
  curves on the top diagram depict the energies of the bottom of the
  conduction band and the top of the valence band as a function of
  position along the dashed line on the cartoon below. The curve
  splits into two inside the ribbon because of the confinement
  gap. The ``transport gap'' can be identified as the amplitude of the
  disorder plus the confinement gap.} 
\label{fig:qdmodel}
\end{figure}

In the latter picture, in which potential inhomogeneities create
a serial arrangement of quantum dots, we expect two distinct ``gaps.''
First, the quantum dot behavior is only apparent when the Fermi level
is close enough to the charge neutrality point that the carrier
density varies spatially from electron-like to hole-like (see Figure
\ref{fig:qdmodel}), 
since otherwise there will be no tunnel barriers between puddles to
form quantum dots. Thus the transport gap (the region of suppressed
conductance at zero bias) is given approximately by the
disorder amplitude plus the confinement gap. The second gap is the
``source-drain gap,'' 
which is roughly the largest value of source-drain voltage for which
conductance is suppressed at some $E_F$. In the simplest case of
single-dot transport, the source-drain gap is the charging energy of
the dot, which need not have a clear dependence on the disorder
amplitude. In the case of multiple quantum dots, determining the
source-drain gap is more complicated, but we still expect the
source-drain gap to depend on the particular shape of the disorder
potential, and the shape of the disorder potential is not strongly
constrained by its amplitude.

In this work, we present transport measurements on graphene 
nanoribbons of 30 nm width and lengths between
30 nm and 3 microns; we highlight several key features of our data
that support the model of
transport through quantum dots produced by charged impurities in the
vicinity of the ribbon. First, by considering the effect of annealing 
nanoribbons to remove impurities from the surface of the ribbon, we
show that the source-drain and transport gaps
cannot be predicted on the basis of geometry alone: the gap 
properties appear to strongly depend on the particular arrangement of
nearby impurities. Second, we show that the
transport gap varies independently from the source-drain gap, in
contrast with the findings of previous work.\cite{Molitor2009}
The transport gap decreases with annealing, which we expect within the
disorder-potential-induced quantum dot model since
annealing should reduce the amplitude of the disorder
potential. Third, similarly to 
the findings of studies on extended graphene sheets,\cite{Chen2008} we find a
connection between the transport gap size and its distance from zero
volts in back gate, suggesting that the transport gap is a measure of
the doping of the sample; the doping is likely connected to the
disorder amplitude. Fourth, we provide data on the relationship between
source-drain gap and ribbon length: we find that in general longer
ribbons have larger gaps, although the scatter in these data is
substantial. Finally, we present measurements on a
particular ribbon sample exhibiting very periodic conductance
oscillations that suggest that the resonances commonly
observed in nanoribbon measurements 
result from Coulomb blockade effects rather than Anderson localization
effects.

\begin{figure}
\centering
\includegraphics[width=3.3in]{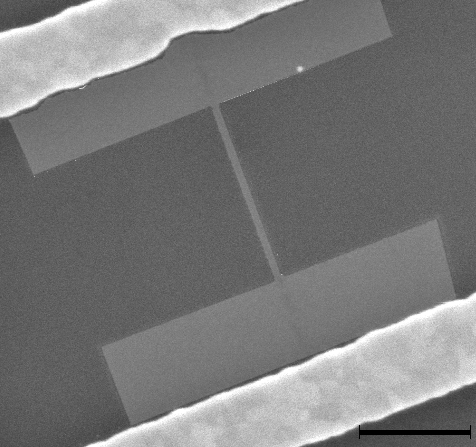}
\caption{Electron micrograph in false color of a typical device, of length
  900 nm and width 30 nm (scale
  bar: 500 nm). Two 
  metal leads (top, bottom) are connected to graphene leads that
  ultimately contact the nanoribbon. A faint dark line can be seen
  extending from the ribbon into the leads; this is a small
  amount of
  residue from the titanium mask.\cite{note1} Image
  taken at 5.0 kV on a 
  FEI XL30 Sirion SEM.}
\label{fig:device}
\end{figure}

Our samples were fabricated with a metal etch mask technique, which
allowed us
to produce nanoribbons with an exposed surface for annealing
experiments as well as a width that is consistent across samples (we
estimate that each ribbon is of width 30 +/- 4 nm, with edge roughness
less than 4 nm; the error bars are determined primarily by the
resolution of 
the SEM used for imaging). First, graphene flakes were deposited on
top of a 
300-nm-thick layer of dry thermal oxide grown on a highly doped
silicon substrate that serves as a back gate; flakes were identified
optically and verified to be single layer by Raman
spectroscopy.\cite{Ferrari2006} Gold contacts with a titanium
sticking layer (15 nm Ti/45 nm Au) were then
patterned to the flake
by electron-beam lithography and electron-beam evaporation. A
30-nm-wide titanium line was patterned 
between contacts to define the ribbon, and a PMMA mask was used to
cover regions near the contacts to preserve graphene
leads at each end of the ribbon;
the whole chip was then 
exposed to oxygen plasma (8 seconds at 65 W) to remove unmasked
graphene. The 
titanium etch mask was washed away in a solution of 30\%
hydrochloric acid at 85$^\circ$C, and the PMMA mask on the leads was removed in
acetone (the HCl penetrates under the PMMA mask so that parts of the titanium
mask still covered by the PMMA are also removed). The resulting devices
consist of metal contacts  
connected to graphene leads that in turn contact the
nanoribbon. A typical device is shown in Figure
\ref{fig:device}.

Measurements were conducted in a cryostat in vacuum at 4.2-4.4 K or while
immersed in liquid helium at 4.2 K unless stated otherwise. For ease of
comparison to other work, we calculate transport gaps
by the method introduced by Molitor et
al.,\cite{Molitor2009} which involves fitting a line to the
regions where conductance increases approximately linearly
surrounding the region where the conductance reaches
zero, and then finding the points where the fit extrapolates to zero
conductance. We 
identify the ``source-drain gap'' by first smoothing the 
data using a 0.5 V window in back gate voltage. We then identify the
source-drain gap as the
largest source-drain voltage below which, for both
positive and negative biases, the (smoothed) differential resistance exceeds
5 M$\Omega$ for some back gate voltage. The smoothing step is
included because there are often one or two outlying ``spike''
features for which the resistance exceeds 5 M$\Omega$ over a
range of source-drain voltages much larger 
than the typical ``gap'' source-drain voltage near the charge
neutrality point; these features have a very small width in
back gate voltage ($\sim 0.1$ V) and are washed out by the
smoothing. While it is possible to choose different and equally valid 
definitions of the transport and source-drain gaps, any one definition
applied consistently across our data sets allows us to make meaningful
statements about the variation of these quantities with ribbon
geometry. Like Molitor et al.,\cite{Molitor2009} we find that the
precise gap definitions used  
do not change the general features of the results.

\begin{table}
\centering
\begin{tabular}{|l|l|}
\hline
Data set & Annealing history \\
\hline
A & Not annealed; data taken while \\ 
& immersed in liquid helium \\
\hline
B & Not annealed; data taken while \\ 
& immersed in liquid helium \\
\hline
C1 & Not annealed; data taken while \\ 
& immersed in liquid helium \\
\hline
C2 & Removed from cryostat after C1, \\
& argon annealed at 300$^\circ$C,\\
 & exposed to atmosphere, cooled down;\\
& data taken in vacuum \\
\hline
C3 & Warmed to RT after C2, current \\
& annealed in vacuum at $\sim 3 \times 10^8$ A/cm$^2$, \\
& cooled down (without breaking vacuum \\
& since C2) \\
\hline
D1 & Not annealed; data taken in vacuum \\
\hline
D2 & Exposed to atmosphere after D1, \\
& current annealed in vacuum at $\sim 5 \times 10^7$\\
& A/cm$^2$, cooled down; data taken in vacuum \\
\hline
D3 & Warmed to RT after D2, current annealed\\
& in vacuum at $\sim 3 \times 10^8$ A/cm$^2$,  cooled down \\
& (without breaking vacuum since D2) \\
\hline
E & Not annealed; data taken in vacuum \\
\hline
\end{tabular}
\caption{Description of data set labels for the samples considered in
  this work. Data sets are named with a letter and sometimes a
  number; sets of the same letter but with different numbers
  correspond to the same group of ribbons on successive cooldowns,
  following successive anneals. Note that each data set can contain
  multiple ribbons. Current densities are calculated assuming a sheet
  thickness of 0.35 nm, consistent with previous work.\cite{Moser2007}}
\label{table:sets}
\end{table}

\begin{figure}
\centering
\includegraphics[width=3.3in]{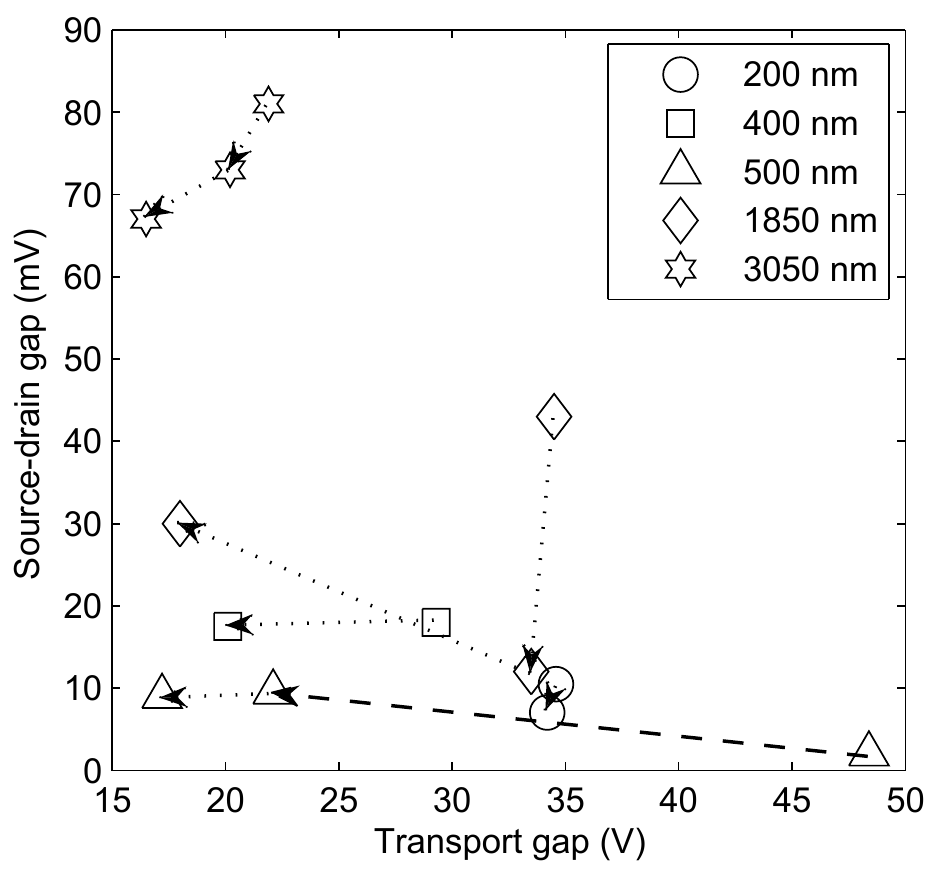}
\caption{Transport gap versus source-drain gap at 4.2 K for a
  representative collection of annealed ribbons of various lengths
  (all 30 nm wide). The 
  arrows point from the 
  gap values before annealing
for a particular sample
to the values for the same sample after annealing. 
The lines with short dashes indicate current annealing performed in
situ, and the line with long dashes indicates annealing in argon at
300$^\circ$C followed 
by exposure to atmosphere and measurement in a vacuum environment.
In general, the
transport gap shrinks 
with annealing, but the source-drain gap can change in either
direction.}
\label{fig:annealed}
\end{figure}

We collected differential conductance data from several sets of ribbons
before and after annealing. The details of the various data
sets are shown in Table \ref{table:sets}. We used two forms of
annealing that have been reported to remove surface impurities in
extended graphene sheets: annealing of the whole chip in argon at
300$^\circ$C,\cite{Elias2009} and current annealing in vacuum, which
heats an individual ribbon 
by Joule heating.\cite{Moser2007, Bolotin2008} A representative set
of results from our annealing 
experiments is shown in Figure \ref{fig:annealed}. 
The first key
feature of these data is that annealing tends to decrease the transport
gap. We expect annealing both to remove
impurities (such as resist residue or contaminants collected while the
sample was exposed to the atmosphere) on the surface of the 
nanoribbon, and to rearrange impurities in the vicinity of the
ribbon. If there were clusters of charged impurities with a dominant polarity
distributed across the surface of the ribbon, we would
expect the amplitude of the disorder potential created by the
impurities to be reduced upon annealing, since the charge density in
the clusters would be 
reduced. Reducing the disorder amplitude 
should reduce the region in $E_F$ where transport is suppressed (i.e.,
reduce the transport gap); the shrinking transport gap thus agrees
with the model of quantum dots forming in the ribbon as a result of
the disorder potential. However, we note that the shrinking transport
gap can agree with any model in which annealing removes
impurities that cause localization. 

The second important result from Figure \ref{fig:annealed} is that the
source-drain gap can change arbitrarily upon annealing. These
unpredictable changes make sense within a quantum dot model;
while annealing is expected to reduce the density of nearby charged
impurities, it is also expected to rearrange those impurities that
remain, resulting in a different configuration of charge
puddles than that before annealing. Since the sizes and locations of the new
puddles should be randomly distributed, the unpredictable
changes of the source-drain gap are understandable:
the source-drain gap is strongly dependent on the sizes of the dots
and the coupling between them.

Were the observed conductance suppression caused primarily by
localization due to edge roughness or defects in the lattice,
we would not expect such dramatic changes in behavior upon
annealing. Recent studies,\cite{CamposDelgado2009} in which
chemically-grown graphene 
nanoribbons were annealed in argon for 30 minutes, have found
correction of lattice point defects and edge 
reconstruction to occur only at
temperatures near 1500$^\circ$C and above; our argon annealing
experiments were carried out at 
300$^\circ$C. Current annealing studies\cite{Jia2009} on
chemically-grown 
nanoribbons also suggest that temperatures of over 2000$^\circ$C
(perhaps near 2800$^\circ$C) must be achieved to reconstruct edges and
improve the crystallinity of the sample by Joule heating. Since
the SiO$_2$ substrate that our ribbons sit on melts around
1650$^\circ$C, and since AFM and SEM measurements reveal no melting of
the substrate after annealing, our ribbons likely do not reach
the temperatures required to cause crystal reconstruction.

\begin{figure*}
\centering
\includegraphics[width=6in]{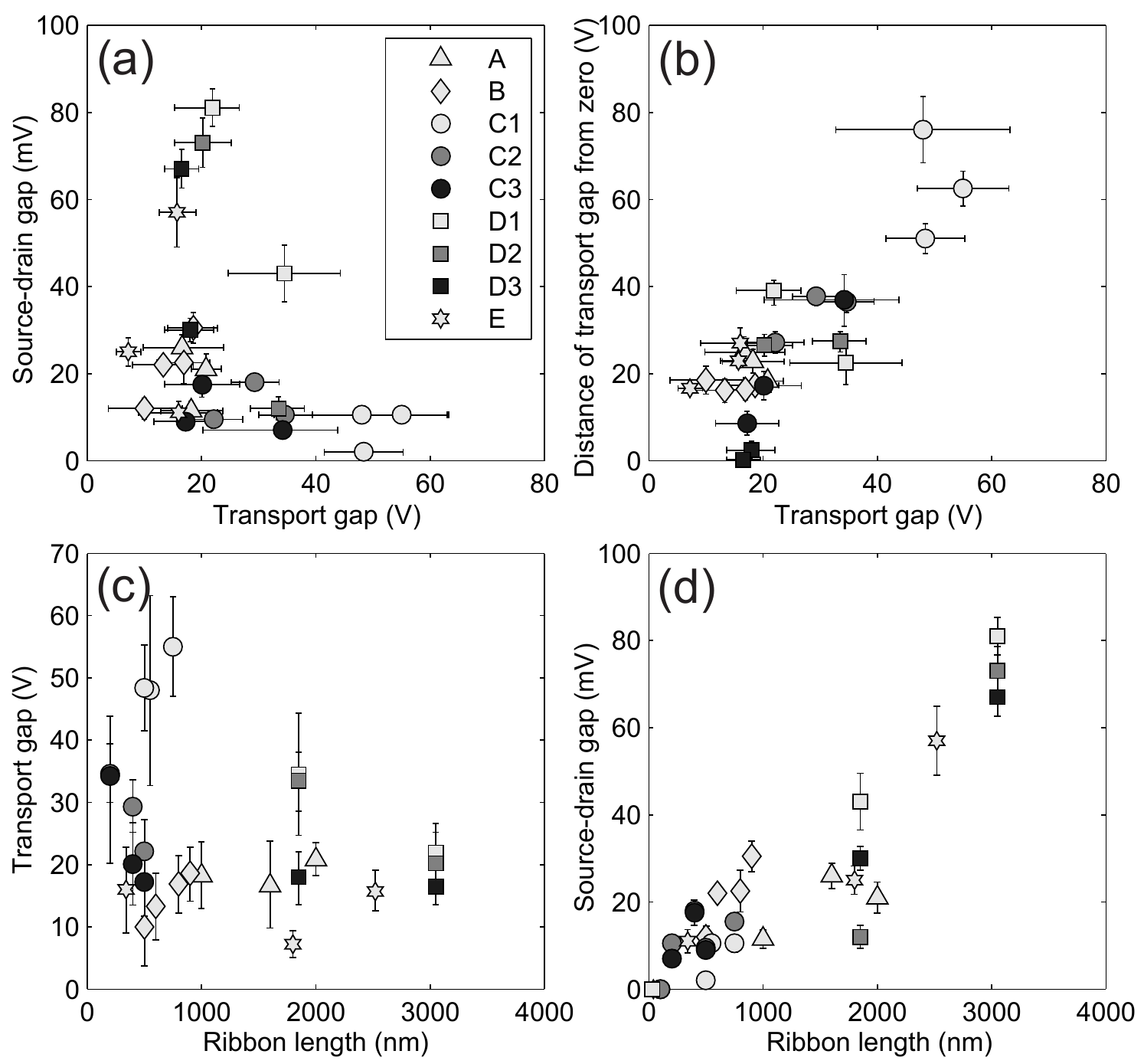}
\caption{ (a) Transport gap versus 
  source-drain gap for all device 
  sets. No clear relationship can be extracted between the source-drain
  and transport gaps, suggesting that the two are independent
  quantities. 
  (b) Transport gap size
  versus distance from 
  zero. As the transport gap increases, its distance
  from zero also typically increases, implying that the transport gap
  is also a 
  measure of sample doping by charged impurities. (c) Ribbon length
  versus transport gap; no connection is observed between length and
  transport gap. Note that the three very disordered outliers from
  data set C1 revert to more common transport gap sizes upon
  annealing. (d) Ribbon length
  versus source-drain gap for all device 
  sets. The source-drain gap generally grows with ribbon length, but
  there is a large amount of scatter, much of it resulting from
  different measurements of the same sample after different annealing
  iterations. 1$\sigma$
  error bars are derived from the linear fits defining the extent of the transport gaps, and from the uncertainty of the mean in our smoothing procedure for the source-drain gap.}
\label{fig:sets}
\end{figure*}

It has been suggested by Molitor et al. that the
source-drain and 
transport gaps are related: in their six samples (widths 30 to 100 nm,
lengths 100 to 500 nm) they noticed an apparently linear 
growth of the transport gap with the source-drain
gap.\cite{Molitor2009} But in our experiments annealing can either
increase or decrease 
the source-drain gap while generally shrinking the transport gap, so
we conclude that
these two gaps are not so simply related. In Figure
\ref{fig:sets}a, we show the transport gap versus
source-drain gap for our samples, in which there is no linear
relationship between gaps. While it is possible that the two data sets
reflect the same underlying physics, it is also possible that
different physics may determine the behavior of their ribbons which
are on average wider and are created via a different process, which
could lead to a different degree of disorder. The lack of a simple
relationship between the source-drain and transport gaps can be
explained within 
the impurity-induced quantum dot picture: the source-drain gap is
sensitive to the specifics of the potential landscape, but the
disorder amplitude, which is roughly identified with the transport
gap, does not strongly constrain the shapes and sizes of the puddles
induced by the potential landscape.

In support of the suggestion that the transport gap is a measure of
the amplitude of the disorder potential is our finding that,
in addition to reducing the size of the transport gap,
annealing tends to shift the center of the transport gap closer to
zero volts in back gate. Assuming that one sign of charged impurity
dominates, the distance 
of the charge neutrality point from zero gate voltage grows with the
number of 
charged impurities, since these impurities dope the sample. Although
this 
assumption need not be valid, in Figure \ref{fig:sets}b we find a
common trend among our samples (annealed or not)  
that the magnitude of the transport gap increases as the distance of
the center of the 
gap from zero in back gate voltage increases. This finding is
reminiscent of the behavior of extended graphene samples upon dosing
with charged impurities; experiment
\cite{Chen2008} and theory \cite{Adam2007} indicate that the width of
the conductivity minimum increases with the distance of the charge
neutrality point from zero gate voltage because transport behavior in
extended graphene sheets is governed by
charged impurity scattering. While there remains some disagreement
about the main scattering mechanism in graphene, our results are
consistent with the charged impurity scattering model. We thus propose
that in our samples
there is a dominant sign of charged adsorbate before annealing
(typically 
negative, since the charge neutrality point is usually at a
positive gate voltage), and that the shifting of  
the charge neutrality point toward zero volts upon annealing results
from the removal of charged impurities. 

As shown in Figure \ref{fig:sets}c, the transport gap has 
little inherent length dependence, consistent with a simple link
between transport gap and disorder amplitude. Certainly, for a short enough
ribbon, there 
is no measurable gap, and our definition of the transport gap is
meaningless. However, this was only the case for ribbons less
than 200 nm long; for longer ribbons, the transport gap as defined
does seem to primarily measure the disorder and not the ribbon length.

Although the transport gap evidently has little length dependence, 
the source-drain gap follows the general trend of increasing as a
function of ribbon length (Figure
\ref{fig:sets}d). The scatter in the data and limited number of data
points prevent extraction of a quantitative trend;
however, the scatter itself is consistent with the impurity
disorder potential picture, in that the gap properties are sensitive to the
particular 
potential profile in the ribbon. Importantly, much of the scatter is
coming from different measurements on the same ribbon after different
annealing procedures, 
so that slight lithographic differences between ribbons (e.g.,
different widths or edge details) cannot be primarily
responsible for the scatter. We also point out that, on average, a longer
ribbon would have more quantum dots whose energies need to
be appropriately positioned with respect to the energies of
their neighbors to enable transport; this would, on average, increase
the bias voltage required to push 
electrons across the ribbon (i.e., increase the source-drain gap).

\begin{figure*}
\centering
\includegraphics[width=6in]{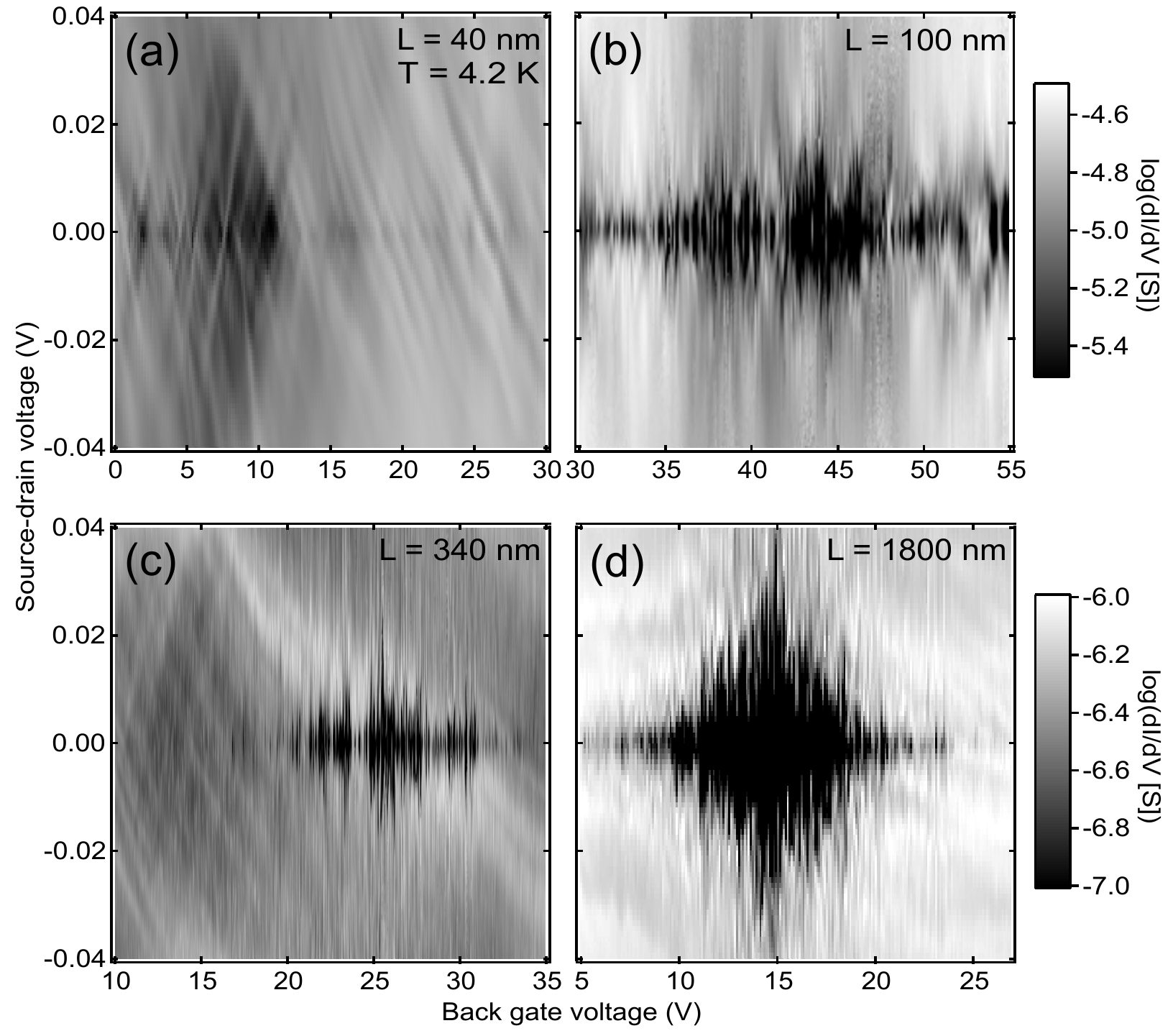}
\caption{Differential conductance versus source-drain and back gate
  voltages for ribbons of length 40 nm (a), 100 nm (b), 340 nm (c),
  and 1800 nm (d). After our smoothing procedure (smoothed data not
shown), both the 40 and 100 nm ribbons have ``zero'' source-drain gap,
while the 340 nm 
ribbon has a nonzero source-drain gap. However, we see some hints of
the typical diamond-shaped gap behavior at a few 
values of gate voltage in (b). Additionally, in (a) and (c),
several intersecting diagonal lines of high conductance can be seen in 
more heavily doped regions of gate voltage. The widths of the resulting
``diamond'' shapes are several volts in back gate. Their geometry is
reminiscent of the Fabry-Perot resonances observed in carbon
nanotubes,\cite{Liang2001} and both their size and geometry 
differ substantially from what we take to be Coulomb diamonds near
the charge 
neutrality point (for a closer look at the Coulomb diamonds, see
Figure \ref{fig:cb_bias_bg} below). We have observed this type of
Fabry-Perot behavior in ribbons that are microns long, as well as in
ribbons that are tens of nanometers long.}  
\label{fig:bias_bg}
\end{figure*}

Another feature in Figure \ref{fig:sets}d is that our 30 and 40
nm-long ribbons had no regions of back gate voltage where conductance
was low enough to be considered ``gapped'' per our definitions. On the
other hand, our ribbons that were at least 100 nm long all had some
gapped regions in gate voltage. Data illustrating the evolution of
the gap for small ribbon lengths are shown in Figure 
\ref{fig:bias_bg}.  If transport is controlled by puddles of
localized carriers surrounded by regions of zero carrier density, we
expect there to be some length below which a puddle is too
well-coupled to the leads to cause strong Coulomb blockade, and thus
there will be no gap. Recent STM experiments indicate that charge
puddles in extended graphene sheets at the charge-neutrality point have an
average length-scale of about 20 
nm. \cite{Zhang2009} Computational studies also suggest that the
typical puddle size in graphene at the charge-neutrality point is
on the order of 10 nm. \cite{Rossi2008} Although we do not know
how this puddle size would be different 
in the case of nanoribbons, the largest ribbon length below which
there is no gap is on the order of tens of nanometers
for these 30-nm-wide 
ribbons; the lengthscales are thus consistent with the
results of the STM experiments and simulations.

\begin{figure*}
\centering
\includegraphics[width=6in]{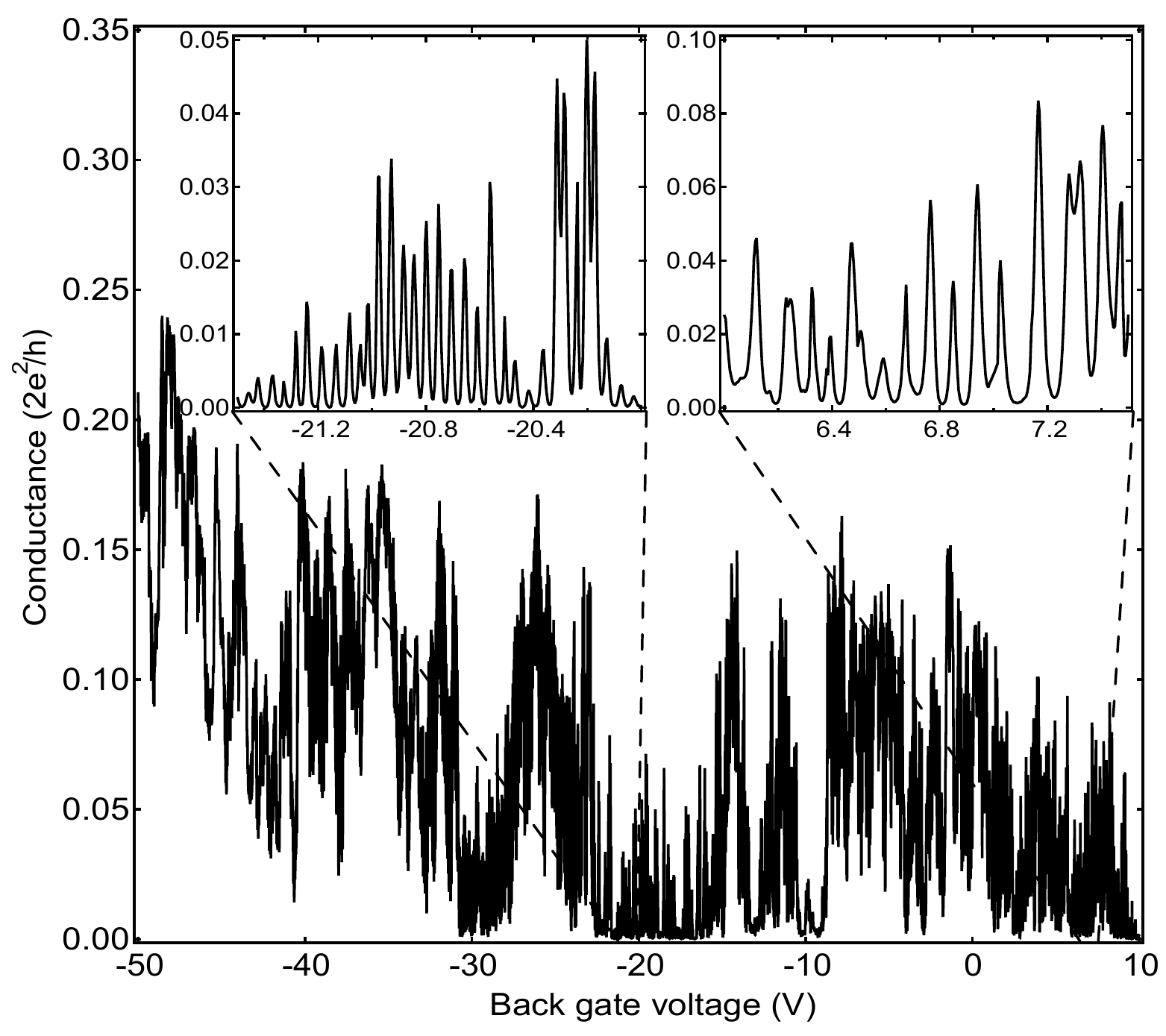}
\caption{Zero-bias conductance versus back gate voltage at 4.2 K of
  a 200-nm-long ribbon displaying highly periodic 
  behavior after annealing. The individual peaks maintain a 
  constant peak spacing over much of the voltage range shown. 
 Left inset: higher-resolution view of
the conductance peaks between -21.5 V and -20 V. The peak
spacing here is 
representative of the peak spacing for the back gate voltage
range -40 V to 5 V. Right inset: higher-resolution view of the
conductance peaks  
for more positive back gate voltages; in this region, the peaks are far
less regular than those shown in the left inset.}
\label{fig:pcb_zerobias}
\end{figure*}

Along with the above aggregate results in support of the
impurity disorder potential model, we 
have observed transport features in one particular ribbon that, after
annealing, strongly resemble Coulomb blockade through one
main quantum dot that may span much of the ribbon's area. In Figure
\ref{fig:pcb_zerobias}, we show a plot of 
conductance versus gate voltage from this sample. In
the left inset of Figure \ref{fig:pcb_zerobias}, we show a closeup of
a small range in 
back gate voltage from the same sweep; on this scale, very regular
conductance peaks can be identified. Hundreds of such peaks occur over
a range 
of more than 40 V in back gate with constant peak spacing. The peaks
survive with the same 
periodicity on the rising background in the more heavily doped
region of negative back gate voltage. As the back gate voltage is increased 
to about 15 V, the peaks become
less periodic. The right inset of Figure
\ref{fig:pcb_zerobias} shows a
closeup of part of the region in gate voltage where the behavior
loses periodicity; while clusters of peaks can have the same
peak spacing, there are a number of peaks at irregular
positions.

\begin{figure}
\centering
\includegraphics[width=3.3in]{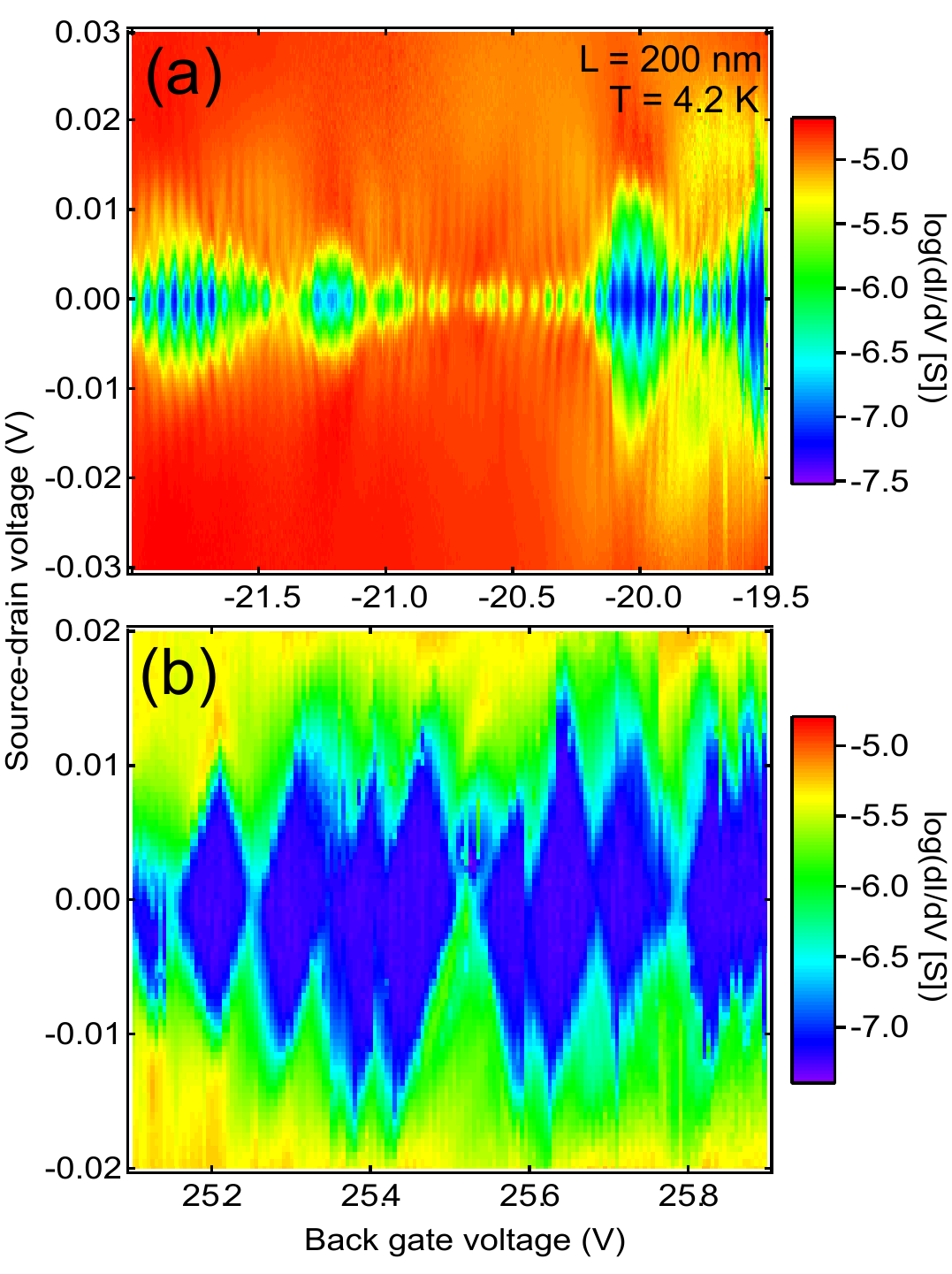}
\caption{(Color online) (a) Differential conductance versus back gate voltage
  and source-drain voltage of a current-annealed 200 nm-long ribbon
  sample at 261 mK. Small, clear diamonds can be observed at a
  frequency of roughly 20 per volt in back gate. The charging energy
  of the dot apparently generating these small diamonds is modulated in
  diamond-like envelopes, which could be Coulomb diamonds arising from
  a small, strongly-coupled dot in series with the larger dot. (b)
  Differential conductance data at 4.2 K 
  of the same device as in (a), but in the region of $E_F$ with less
  periodic peaks. Here, overlapping diamond features, like those
  expected for serial Coulomb blockade, can be seen. 
  (A parallel arrangement of quantum dots could also look like
  overlapping diamonds, but the conductance peaks of the
  overlapping diamonds would be present in the overlap region; in a
  serial configuration, the overlap regions lack conductance
  peaks, since electrons must tunnel through both dots to contribute
  to conductance.\cite{vanderWiel2002})
}
\label{fig:cb_bias_bg}
\end{figure}

In plots of differential
conductance versus bias and back gate voltage in the periodic region,
such as that in Figure  
\ref{fig:cb_bias_bg}a, features that resemble the
Coulomb diamonds of a single quantum dot are apparent. The heights
of these diamonds are  
modulated as a function of gate voltage such that adjacent diamonds
form diamond-like packets,
suggesting the presence of 
another physically smaller quantum dot that is strongly coupled to a
lead and is 
acting in series with the larger quantum dot responsible
for the smaller 
diamonds. In contrast, Figure \ref{fig:cb_bias_bg}b shows 
the less periodic region, in which we find ``overlapping'' diamond
features characteristic of Coulomb blockade effects in a serial
arrangement of multiple weakly coupled quantum dots. 

\begin{figure}
\centering
\includegraphics[width=3.3in]{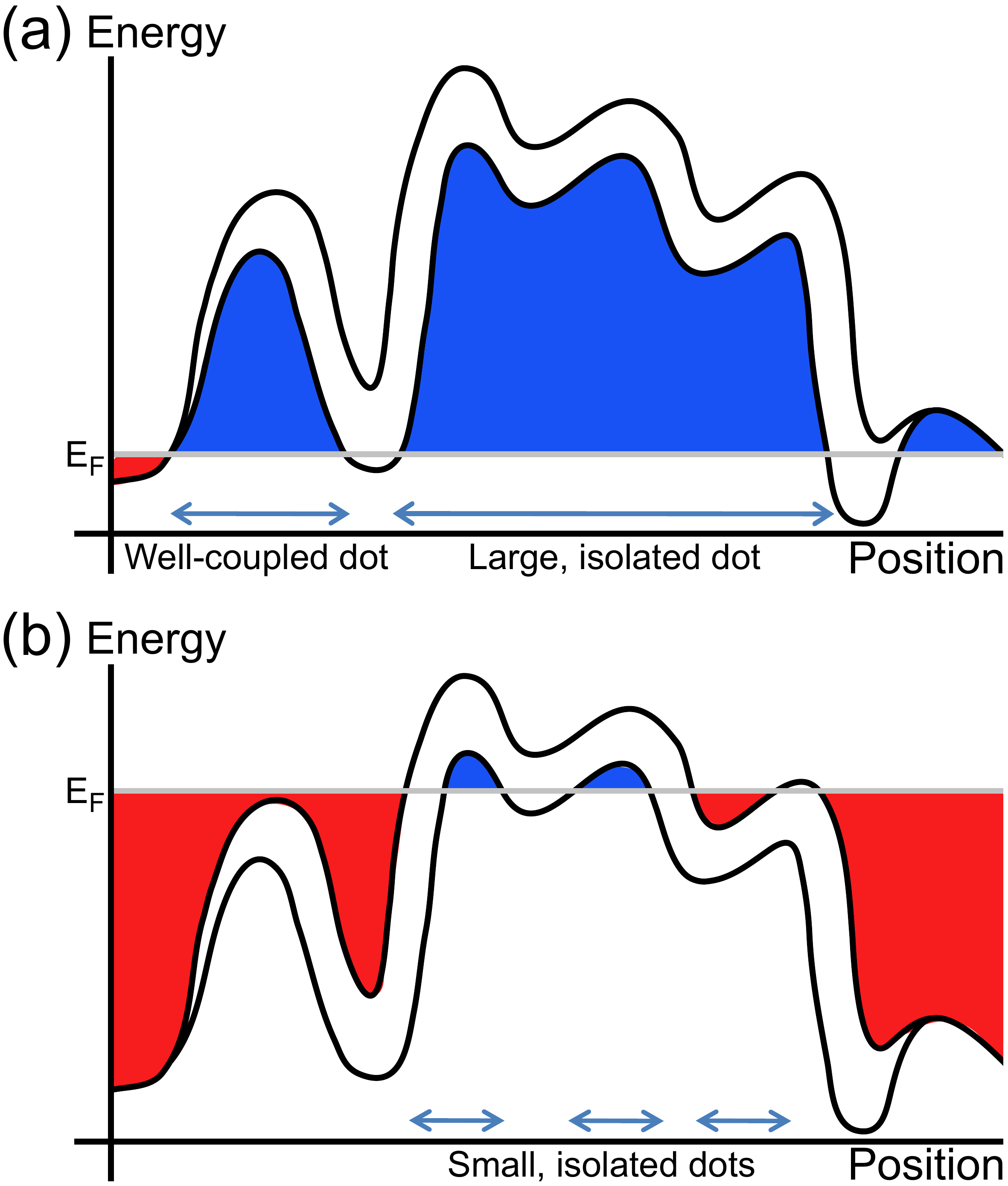}
\caption{(Color online) Cartoon of a potential profile (energy versus
  position along 
  the ribbon) that could give rise to the behavior observed in our
  sample. (a) The Fermi level is set low enough such that there is
  one isolated dot that spans most of the ribbon, as well as another
  dot that is well-coupled to one lead. This dot configuration could
  lead to the  
  slowly modulated but otherwise highly periodic Coulomb oscillations
  shown in the left inset of Figure \ref{fig:pcb_zerobias}. As the
  Fermi level is 
  raised, the sizes and couplings of the dots change slightly, but
  this double-dot system persists essentially unaltered for a range of
  Fermi levels. (b) Once the Fermi 
  level is high enough, the 
  ribbon splits up into several small, isolated quantum dots in
  series. We expect this system to exhibit aperiodic oscillations like
  those shown in the right inset of \ref{fig:pcb_zerobias}.
}
\label{fig:s_to_m}
\end{figure}

The transition from periodic to less periodic behavior as the Fermi
level is moved through the transport gap is
suggestive of Coulomb blockade occuring due to potential
inhomogeneities. In Figure \ref{fig:s_to_m}a, we show a cartoon
representation of a potential
profile, composed of one large  
``well'' with some weaker modulation inside the well, which could give
rise to the observed phenomena.
For some range of $E_F$ that includes the $E_F$ shown in Figure
\ref{fig:s_to_m}a, there is only one isolated puddle, which spans 
most of the 
ribbon. There is also a smaller puddle that is strongly coupled to one
lead. But when $E_F$ is in the range of the weaker
modulation inside the well (Figure \ref{fig:s_to_m}b) the
ribbon splits up into more than one 
isolated puddle; this is a serial arrangement of multiple weakly
tunnel-coupled quantum dots.

We note that it is possible that the larger, more isolated quantum dot
that exists in the  
periodic region of gate voltage does not span as much of the ribbon as
we have shown 
in Figure \ref{fig:s_to_m}a. However, if the dot takes up only a fraction
of the ribbon, the
potential must be exceptionally smooth; any substantial roughness will
create other isolated quantum dots near where the disorder potential
energy
crosses the Fermi level. Since our system apparently only has one
isolated dot, and since an extremely smooth disorder potential seems
unlikely, we propose that the isolated quantum dot does span most of
the ribbon. 
 
To quantitatively estimate the dot size, we use the peak
spacing ($\sim 0.05$ V) along with the capacitance per unit length
calculated by Lin et al.\cite{Lin2008} ($\sim$ 3.5 $\times$ 10\superscript{-20} F/nm)
for a graphene nanoribbon of 
30 nm width above 300 nm of SiO$_2$. We find a dot area of 3000 nm$^2$,
which corresponds to a 100 nm-long dot if the dot spans the width of
the ribbon. This agrees with our qualitative argument that the main
dot covers most of the ribbon. 

We think it unlikely that
the dot in this ribbon is nucleated by a single impurity. Based on the
mobilities of our devices measured outside the ribbon body, following earlier models \cite{Chen2008} we estimate the density of charged
impurites near the graphene sheet to be $\sim 2 \times 10^{13}$
$\textrm{cm}^{-2}$ before annealing. Studies \cite{Chen20082} on
extended graphene flakes have suggested that the maximum achieveable 
mobilities for graphene on SiO$_{\textrm{2}}$ are limited to  $\sim$ 15,000 cm$^{\textrm{2}}$/Vs, corresponding to
an impurity density of roughly $1 \times 10^{11}$
$\textrm{cm}^{-2}$. Even if we were to achieve this maximum mobility
after several annealing attempts, we would still expect more than one
impurity within the area of this ribbon. Since clustering of
impurities and overscreening effects are likely, we do not
expect a one to one correspondence between impurities and charge
puddles. 

Importantly, the gradual transition from periodic to aperiodic
behavior observed in Figures \ref{fig:pcb_zerobias} and \ref{fig:cb_bias_bg}
suggests  
that the less periodic peaks are also Coulomb blockade peaks.
These conductance peaks in the less periodic region are
similar in width and lineshape to those in our other ribbons that
exhibit little periodic behavior; given this point, 
Coulomb blockade phenomena are likely being observed in our other ribbon
samples as well. While it has been proposed that
Anderson localization can create conductance peaks and transport gap
phenomena,\cite{Mucciolo2009} peaks created by Anderson localization would
not occur with highly periodic spacing over a wide range of Fermi energies. 

In summary, we have provided evidence in support of a model of
nanoribbon behavior in which charged impurities in the vicinity of the
ribbon create a
disorder potential that, coupled with some small energy gap, breaks the
ribbon up into isolated puddles of charge carriers that act as quantum
dots. By performing annealing studies on our nanoribbons, we have
demonstrated that the source-drain and transport gaps are distinct
quantities, as expected within the quantum dot model. Our data further
imply that the transport gap reflects 
the doping of the sample as well as its
disorder amplitude, suggesting that the gap phenomena
arise in large part from disorder due to charged impurities.  We
showed that 
the source-drain gap is not a simple function of ribbon length and
width, but that it seems to depend sensitively on the potential
profile in the nanoribbon; this result is understandable within the
quantum dot framework since the dots' sizes, positions, and tunnel
barriers are controlled by the precise potential profile, and the
transport properties of a system of quantum dots depend heavily on
these parameters. Nonetheless, longer ribbons tend to have
larger source-drain gaps, and very short ribbons exhibit no gap
behavior; this behavior is understood within our model, since there is
some smallest length required to fit a well-isolated quantum dot in
the ribbon to block transport, and since a longer ribbon will on
average have
more quantum dots that electrons must tunnel across for conduction,
increasing the source-drain gap. Finally, we provided data from a 
200 nm long ribbon 
displaying highly periodic modulations of conductance versus gate
voltage; we have 
identified these modulations as Coulomb blockade oscillations based
on their similarity to Coulomb oscillations in single-quantum dot
systems. This 
supports the idea that the conductance peaks commonly observed in
nanoribbons arise from Coulomb blockade rather than from Anderson 
localization due to edge disorder.

The apparent importance of
disorder in determining transport properties of lithographically-defined
graphene nanoribbons raises questions
about the feasibility of using such ribbons as next-generation
transistor technology. 
We acknowledge that the influence of disorder due to charged
impurities near the ribbon relative to the influence of atomic-scale
disorder at the ribbon edges may differ in ribbons fabricated by
different techniques. This variation may occur even in ribbons
lithographically fabricated via differing masking techniques. More
radically, new methods have been proposed for making nanoribbons with
atomically ordered edges.\cite{Li2008, Jiao2009}  However, we believe that our
results demonstrate that 
unless the strength of disorder due to charged impurities near the
ribbon can also be reduced
(for instance, 
by suspending\cite{Bolotin2008} the nanoribbons), the
transport behavior of these clean-edged nanoribbons will still be
dominated by the Coulomb blockade of multiple quantum dots.
But perhaps, by carefully controlling the
amount of disorder in these ribbons, Coulomb blockade effects could be
harnessed to create reliable switching behavior. 

\begin{acknowledgements}
We thank Gil Refael, Antonio Castro Neto, Eduardo Mucciolo, Misha
Fogler, Enrico Rossi, Shaffique Adam, Stephan Roche, Sami Amasha, and Jimmy Williams
for very useful discussions. We also thank Emerson Glassey for
assistance with sample fabrication. 
This work was supported by the Center for Probing the Nanoscale, an
NSF Nanoscale Science and Engineering Center, 
award PHY-0830228, through a Supplementary
award from the NSF and the NRI; as well as by the Focus Center
Research Program's Center on Functional Engineered Nano Architectonics
(FENA). P. Gallagher acknowledges support
from the Stanford Vice Provost for Undergraduate Education. K. Todd
acknowledges support from the Intel and Hertz
Foundations. D. Goldhaber-Gordon acknowledges support from Stanford
under the Hellman Faculty Scholar program. Work
was performed in part at the Stanford Nanofabrication Facility (a
member of the National Nanotechnology Infrastructure Network), which is
supported by the National Science Foundation under Grant ECS-9731293,
its lab members, and the industrial members of the Stanford Center for
Integrated Systems.  
\end{acknowledgements}


\begin{thebibliography}{10}

\bibitem{CastroNeto2009}
A.~H. {Castro Neto} {\it et~al.}, Rev. Mod. Phys. {\bf 81},  109  (2009).

\bibitem{Han2007}
M.~Y. {Han}, B. {{\"O}zyilmaz}, Y. {Zhang}, and P. {Kim}, Phys. Rev. Lett. {\bf
  98},  206805  (2007).

\bibitem{Chen2007}
Z. {Chen}, Y.-M. {Lin}, M.~J. {Rooks}, and P. {Avouris}, Physica E {\bf 40},
  228  (2007).

\bibitem{Brey2006}
L. {Brey} and H.~A. {Fertig}, Phys. Rev. B {\bf 73},  235411  (2006).

\bibitem{Ezawa2006}
M. {Ezawa}, Phys. Rev. B {\bf 73},  045432  (2006).

\bibitem{Yang2007}
L. {Yang} {\it et~al.}, Phys. Rev. Lett. {\bf 99},  186801  (2007).

\bibitem{Todd2009}
K. {Todd}, H.-T. {Chou}, S. {Amasha}, and D. {Goldhaber-Gordon}, Nano Lett.
  {\bf 9},  416  (2009).

\bibitem{Molitor2009}
F. {Molitor} {\it et~al.}, Phys. Rev. B {\bf 79},  075426  (2009).

\bibitem{Evaldsson2008}
M. {Evaldsson}, I.~V. {Zozoulenko}, H. {Xu}, and T. {Heinzel}, Phys. Rev. B
  {\bf 78},  161407(R)  (2008).

\bibitem{Mucciolo2009}
E.~R. {Mucciolo}, A.~H. {Castro Neto}, and C.~H. {Lewenkopf}, Phys. Rev. B {\bf
  79},  075407  (2009).

\bibitem{Stampfer2009}
C. {Stampfer} {\it et~al.}, Phys. Rev. Lett. {\bf 102},  056403  (2009).

\bibitem{Liu2009}
X. {Liu}, J.~B. {Oostinga}, A.~F. {Morpurgo}, and L.~M.~K. {Vandersypen}, Phys.
  Rev. B {\bf 80},  121407(R)  (2009).

\bibitem{Sols2007}
F. {Sols}, F. {Guinea}, and A.~H. {Castro Neto}, Phys. Rev. Lett. {\bf 99},  166803
   (2007).

\bibitem{Nomura2007}
K. {Nomura} and A.~H. {MacDonald}, Phys. Rev. Lett. {\bf 98},  076602  (2007).

\bibitem{Nomura20072}
K. {Nomura}, M. {Koshino}, and S. {Ryu}, Phys. Rev. Lett. {\bf 99},  146806
  (2007).

\bibitem{Bardarson2007}
J.~H. {Bardarson}, J. {Tworzyd{\l}o}, P.~W. {Brouwer}, and C.~W.~J.
  {Beenakker}, Phys. Rev. Lett. {\bf 99},  106801  (2007).

\bibitem{Chen2008}
J.-H. {Chen} {\it et~al.}, Nat. Phys. {\bf 4},  377  (2008).

\bibitem{note1}
We consider the question of whether all traces of the titanium mask are removed
  in the acid solution. It appears that residual titanium is not affecting
  transport properties, since the ribbons made by this process behave similarly
  to those made by other processes. Additionally, we prepared a test bulk
  sample by coating part of it in titanium and then etching away the titanium
  using the same process as was used for the nanoribbons. Four-wire conductance
  measurements revealed characteristic graphene behavior before the titanium
  etch mask was deposited and after it was removed, with a mobility degradation
  of the degree expected from additional lithography: the mobility decreased
  from 6000 cm$^2$V$^{-1}$s$^{-1}$ to 800 cm$^2$V$^{-1}$s$^{-1}$ after titanium
  deposition and removal. A bulk mobility of 800 cm$^2$V$^{-1}$s$^{-1}$ is
  typical for the nanoribbon devices that we have fabricated in the past using
  two steps of lithography with no titanium deposition or titanium etch.\cite
  {Todd2009} Raman spectroscopy revealed no changes in the disorder peak after
  the titanium deposition and removal, and the Raman spectrum did not reveal
  any differences between the parts of the flake that were once covered with
  titanium and the parts that were not covered with titanium. Auger
  measurements from a PHI 700 found no residual titanium to within the
  sensitivity of the instrument: the titanium signal was the same on all parts
  of the devices studied, regardless of whether or not titanium had been
  deposited there to begin with. Based on Auger measurements of a titanium film
  of known thickness, we estimate that any residual titanium film on our
  nanoribbons is less than 0.2 nm thick. While this upper bound is rather
  large, based on all the evidence presented here, we feel that the titanium
  mask is being sufficiently removed by our etch process and that the amount of
  residual titanium on our ribbons is much smaller than this upper bound.

\bibitem{Ferrari2006}
A.~C. {Ferrari} {\it et~al.}, Phys. Rev. Lett. {\bf 97},  187401  (2006).

\bibitem{Moser2007}
J. {Moser}, A. {Barreiro}, and A. {Bachtold}, App. Phys. Lett. {\bf 91},
  163513  (2007).

\bibitem{Elias2009}
D.~C. Elias {\it et~al.}, Science {\bf 323},  610  (2009).

\bibitem{Bolotin2008}
K.~I. {Bolotin} {\it et~al.}, Solid State Commun. {\bf 146},  351  (2008).

\bibitem{CamposDelgado2009}
J. Campos-Delgado {\it et~al.}, Chem. Phys. Lett. {\bf 469},  177  (2009).

\bibitem{Jia2009}
X. Jia {\it et~al.}, Science {\bf 323},  1701  (2009).

\bibitem{Adam2007}
S. {Adam}, E.~H. {Hwang}, V.~M. {Galitski}, and S. {Das Sarma}, PNAS {\bf 104},
   18392  (2007).

\bibitem{Liang2001}
W. Liang {\it et~al.}, Nature {\bf 411},  665  (2001).

\bibitem{Zhang2009}
Y. {Zhang} {\it et~al.}, Nat. Phys. {\bf 5},  722  (2009).

\bibitem{Rossi2008}
E. {Rossi} and S. {Das Sarma}, Phys. Rev. Lett. {\bf 101},  166803  (2008).

\bibitem{vanderWiel2002}
W.~G. {van der Wiel} {\it et~al.}, Rev. of Mod. Phys. {\bf 75},  1  (2002).

\bibitem{Lin2008}
Y.-M. {Lin}, V. {Perebeinos}, Z. {Chen}, and P. {Avouris}, \prb {\bf 78},
  161409(R)  (2008).

\bibitem{Chen20082}
J.-H. {Chen} {\it et~al.}, Nat. Nanotechnol. {\bf 3},  206  (2008).

\bibitem{Li2008}
X. {Li} {\it et~al.}, Science {\bf 319},  1229  (2008).

\bibitem{Jiao2009}
L. {Jiao} {\it et~al.}, Nature {\bf 458},  877  (2009).

\end{thebibliography}
\end{document}